\documentclass[twocolumn,prl,showpacs,amsmath]{revtex4}

\usepackage{graphicx}
\usepackage{bm}
\usepackage{epsf}

\newcommand{\beq}{\begin{equation}}
\newcommand{\eeq}{\end{equation}}
\newcommand{\beqa}{\begin{eqnarray}}
\newcommand{\eeqa}{\end{eqnarray}}
\newcommand{\beqan}{\begin{eqnarray*}}
\newcommand{\eeqan}{\end{eqnarray*}}
\newcommand{\ben}{\begin{enumerate}}
\newcommand{\een}{\end{enumerate}}
\newcommand{\bit}{\begin{itemize}}
\newcommand{\eit}{\end{itemize}}

\begin{document}

\title{Pulse-induced acoustoelectric vibrations in surface-gated GaAs-based quantum devices \\}

\author{S. Rahman$^1$}
\email{S.Rahman.00@cantab.net}
\author{T. M. Stace$^2$}
\author{H. P. Langtangen$^3$}%
\author{M. Kataoka$^1$}%
\author{C. H. W. Barnes$^1$}%
\affiliation{$^1$Cavendish Laboratory, Cambridge University, J J Thomson Avenue, Cambridge, CB3 OHE, United Kingdom}%
\affiliation{$^2$Centre for Quantum Computation, DAMTP, Centre for
Mathematical Sciences, Cambridge University, Wilberforce Road, Cambridge, CB3 0WA, United Kingdom}%
\affiliation{$^3$Simula Research Laboratory, Martin Linges v 17, Fornebu P.O.Box 134, 1325 Lysaker, Norway}%

\date{\today}

\begin{abstract}

We present the results of a numerical investigation which show the
excitation of acoustoelectric modes of vibration in GaAs-based
heterostructures due to sharp nano-second electric-field pulses
applied across surface gates. In particular, we show that the
pulses applied in quantum information processing applications are
capable of exciting acoustoelectric modes of vibration including
surface acoustic modes which propagate for distances greater than
conventional device dimensions. We show that the pulse-induced
acoustoelectric vibrations are capable of inducing significant
undesired perturbations to the evolution of quantum systems.

\end{abstract}


\maketitle

\section*{INTRODUCTION}

Quantum computers have the potential to efficiently solve ``hard
problems'', for which there are no known efficient classical
algorithms \cite{NielsonBook}. The realization of a quantum
processor is currently a major challenge at the forefront of
experimental physics. Semiconductor quantum dots form the basis of
many solid-state devices being developed for the implementation of
quantum logic gates. They are believed to be promising candidates
for producing scalable systems, as they take advantage of the
technology available for the fabrication of semiconductor
technology. The essential requirements for the realization of such
devices are the coherent manipulation of an electronic degree of
freedom in each dot and the control of coupling between adjacent
dots. Since the theoretical work of Loss and DiVincenzo
\cite{Loss:98}, experimental progress has been made in the
realization of spin qubits \cite{Petta:05} and charge qubits
\cite{Hayashi:04} based on quantum dots formed by the surface gating
of a two-dimensional electron gas (2DEG) where voltage pulses are
used to control exchange interaction and tunnelling, respectively,
between adjacent dots. However, the scaling of such systems
requires the identification and minimization of the decoherence
processes, which arise from the coupling of the quantum dot to its
environment. This requires a precise understanding of the dynamical
evolution of both the relevant electronic degree of freedom and the
environment to which it is coupled.

In our recent theoretical work \cite{Rahman:05:1,Rahman:05:2}, we
had implemented mechanical boundary conditions in the form of
time-dependent traction forces on the surface of a GaAs-based
device, to excite acoustoelectric vibrations. By utilizing the
piezoelectric nature of GaAs, time-varying electric fields form the
basis of surface acoustic wave generation through interdigitated
transducers. The time-varying electric fields such as those used to
manipulate spin and charge qubits in quantum information processing
applications may also excite acoustoelectric vibrations. These
vibrations would be undesirable for the operation of sensitive
quantum dot systems.

In this paper, we show through the numerical solution of the
equations of motion in GaAs, the excitation of acoustoelectric
modes of vibration in GaAs-based heterostructure devices, due to
pulsed electric fields. We discuss the results in relation to the
construction of scalable quantum dot devices.

\section*{THEORY}

The equations of motion in a general piezoelectric material are
\cite{Auld}

\begin{equation}\label{eqn:Electrical}
 \frac{\partial }{\partial x_i} \epsilon^S_{ij} \frac{\partial \phi}{\partial x_j}
=  \frac{\partial }{\partial x_i} e_{ijk} \frac{\partial
u_k}{\partial x_j},
\end{equation}

\begin{equation}\label{eqn:Mechanical}
     \varrho \frac{\partial^2u_i}{\partial t^2} =  \frac{\partial}{\partial x_j} c_{ijkl}^E \frac{\partial u_l}{ \partial x_k} +
    \frac{\partial }{\partial x_j} e_{kij}  \frac{\partial \phi}{\partial x_k},
\end{equation}

\noindent where $u_i$ is a displacement vector, $\phi$ is the
electrostatic potential field, $c^{E}_{ijkl}$ are components of
the elastic tensor, $\epsilon^S_{ij}$ are components of the
material dielectric (the superscripts E and S indicate
measurements under constant strain and electric field
respectively), $\varrho$ is the mass density, and $e_{ijk}$ are
components of the piezoelectric tensor.

The mechanical boundary conditions require the prescription of the
traction forces $\sigma_{ij} \cdot \hat{n}$ at each external
surface, where $\hat{n}$ is a unit vector perpendicular to an
external surface, and $\sigma_{ij}$ are components of the stress
tensor, defined by

\begin{equation}\label{eqn:Tij1}
    \sigma_{ij} = -e_{kij} E_k + c^E_{ijkl} \varepsilon_{kl},
\end{equation}

\noindent where $E_k = -\frac{\partial \phi}{\partial x_k}$ and
$\varepsilon_{ij}  = \frac{1}{2} (\frac{\partial u_i}{\partial
x_j} + \frac{\partial u_j}{\partial x_i})$. The electrostatic
boundary conditions require the prescription of the normal
component of the electric displacement at the free surface of the
medium. The electric displacement vector is defined by

\begin{equation}\label{eqb:eBC1}
    D_i = \epsilon_{ij}^S E_j + e_{ijk} \varepsilon_{jk}.
\end{equation}

The equations consist of a system of coupled linear partial
differential equations second order in space and time. The
solution of these equations for non-trivial boundary conditions
requires numerical methods. Our numerical solution method, which
utilizes a finite element approximation for the spatial part and a
second order finite difference discretisation in time, is suitable
for the simulation of electromechanical waves \cite{Rahman:05:1}.
We have used eight-noded brick elements corresponding to linear
basis functions $N_i$, resulting in an overall spatial and
temporal convergence rate of 2 for the error. An operator
splitting strategy is employed to split the coupled $u_i$-$\phi$
problem \cite{DpBook2}. The numerical formulation and verification
is described in more detail in Ref.~\onlinecite{Rahman:05:1}.

\begin{figure}

\epsfxsize=5.25cm

\centerline{\epsffile{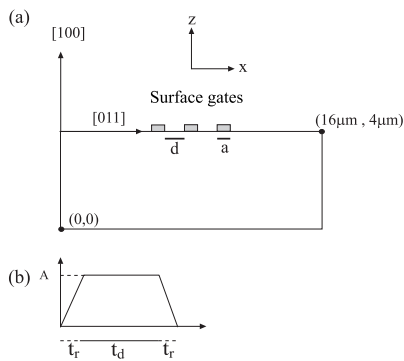}}

\caption{(a) A schematic representation of the device used in our
simulations. The orientation of the crystal axes in relation to the
coordinate axes is shown. The parameters $a$ and $d$ represent the
width of the gates and distance between the gates, respectively. (b)
The shape of the pulse used in the electrical pulse experiments. The
parameters $A$, $t_r$ and $t_d$ are the amplitude, rise time and
pulse duration respectively.}

\label{fig:pulseShape}

\end{figure}

For GaAs, the non-zero components of the piezoelectric tensor have
the value $e_{14} = -0.16$~Cm$^{-2}$. The non-vanishing components
of the permittivity tensor are $\epsilon_{11} = \epsilon_{22} =
\epsilon_{33} = \epsilon_{s} = 13.18 \epsilon_0$ ($\epsilon_0 =
8.85 \cdot 10^{-12}$~Fm$^{-1}$), and the non-vanishing components
of the elastic tensor are $c_{xxxx}$ = $c_{yyyy}$ = $c_{zzzz}$ =
$c_{11} $ = $11.88 \cdot 10^{10}$~Nm$^{-2}$, $c_{xxyy}$ =
$c_{yyzz}$  = $c_{zzxx}$ = $c_{12}$ = $5.38 \cdot
10^{10}$~Nm$^{-2}$, and $c_{xyxy}$ = $c_{yzyz}$ = $c_{zxzx}$ =
$c_{44}$ = $5.94 \cdot 10^{10}$~Nm$^{-2}$. All other non-zero
components of the elasticity tensor can be determined by applying
its symmetry properties $c_{ijkl} = c_{jikl} = c_{ijlk} =
c_{klij}$. The mass density of GaAs $\varrho$ has the value of
$5.36 \cdot 10^3$~Kgm$^{-3}$ \cite{Adachi}.

Figure~\ref{fig:pulseShape}(a) shows a schematic representation of
the system we have modelled in our simulations. The coordinate $x$
and $z$ -axes are aligned with the crystal [011] and [100]
directions, respectively, as is frequently the case in the
realization of GaAs-based quantum dot systems. For this
orientation, the properties of acoustoelectric vibrations are well
documented \cite{Simon,Rahman:05:2} and hence it is also
convenient for the interpretation of our results.

The anisotropic nature of GaAs will result in some dependence of
the acoustic properties on the crystal orientation. However, this
is expected to be small as the components of the piezoelectric
tensor are equal and the components of the elasticity tensor are
the same order of magnitude. In general, quantum dot applications
will include electric fields across components in the orientation
investigated in this paper. The presence of a thin layer of AlGaAs
is neglected for simplicity, and has been shown to have little
effect on the acoustic properties \cite{Rahman:05:2}. As in our
previous work we avoid a Hartree-Fock type self-consistent
calculation as the benefits of this are likely to be small
compared to drawbacks associated with vast computational
resources.

In our simulations, we implement a trapezoidal form for the
applied voltage pulse as shown in Fig.~\ref{fig:pulseShape}(b).
Trapezoidal pulses are implemented in experiments where coherent
manipulation or measurement of single electron charge and spin
-systems have been achieved
\cite{Hayashi:04,Elzerman:04,Petta:05}.
Figure~\ref{fig:pulseShape}(b) shows the geometrical features of
the pulse. In experimental demonstrations, the parameters of the
pulse are such that $ t_r << 0.1~\text{ns}$, $ 0.1~\text{ns} < t_d
< 1000.0~\text{ns}$. We apply the voltage bias to the central gate
relative to the outer gates. The gate width and separation are
chosen to be those frequently used in experiments to define
quantum dots; $a=250$~nm $d=250$~nm. The mechanical gates are
excluded from our simulations as they are capable of perturbing
the induced vibrations \cite{Rahman:05:2}. The presence of
mechanical gates has been shown to reduce the amplitude of surface
acoustic modes but also affect its vibrational frequency. To model
the applied voltage bias, we apply Dirichlet boundary conditions
on the regions of the surface which the gates would occupy and a
Neumann-type condition elsewhere through setting $D_z = 0$
\cite{ft0}. For the mechanical boundary conditions, we apply
traction-free boundary conditions i.e. $\sigma_{ij} \cdot \hat{n}
= 0$ on the external surfaces.

The precise characterization of the dependence of the effects of a
pulse on all the parameters of this system is beyond the scope or
intention of this paper. Our aim is to show the effects on the
underlying lattice of electric-field pulses frequently used in
quantum dot experiments. We first demonstrate the effects of a
``typical'' pulse used in experiments, and then consider the
dependence of these effects on the parameters of the applied pulse,
$t_d$ and $t_r$.

\section*{RESULTS AND DISCUSSIONS}

\begin{figure}[!htb]
\begin{center}
\epsfxsize=8.5cm

\centerline{\epsffile{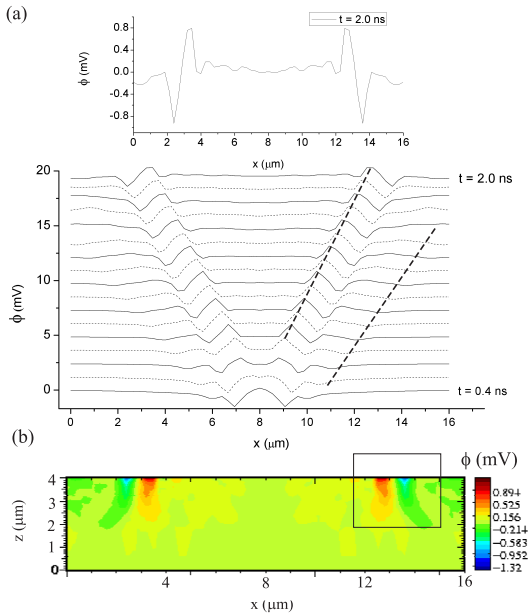}}
\end{center}
\caption{(a) Time series plot of the electric potential as a
function of $x$ coordinate from time $t=0.4$~ns to $t=2.0$~ns, with
$\Delta t = 0.1$~ns. The induced waves propagate outward from the
center at $x = 8~\mu$m with increasing time. For clarity, the inset
shows the curve at $t=2.0$~ns with a maximum amplitude of $\sim
2.0$~mV. The data was taken at a depth of $100$~nm from the surface.
(b) Grayscale plot of the pulse induced electric potential at time
$t = 2.0$~ns. The regions of higher electric field at ($x \sim
3~\mu$m) and ($x \sim 13~\mu$m outlined) correspond to single minima
(or maxima) of a surface acoustic wave.} \label{fig:pulseFirstSim}
\end{figure}

Figure~\ref{fig:pulseFirstSim}(a) shows the oscillations of the
electric potential $\phi$, along the $x$ axis and $100$~nm below
the surface (as this is the standard depth of a 2DEG), resulting
from a pulse with $A = 1$~V, $t_r = 0.025$~ns and $t_d = 0.3$~ns.
From time $t = 0.4$~ns, the waves propagate outward from the
center at velocities ranging from $\sim$ $2750$~ms$^{-1}$ to
$\sim$ $5000$~ms$^{-1}$, with the lower velocities likely to
correspond to surface acoustic waves \cite{Simon} and the higher
velocities to other modes which include bulk waves.

As the inset of Fig.~\ref{fig:pulseFirstSim}(a) shows, the maximum
amplitude of the waves at the time $t = 2.0$~ns is $\sim 2.0$~mV.
The linearity of the underlying equations of motion would require a
linear relationship between the amplitude of the applied pulse, and
that of the pulse induced oscillations. The pulse induced vibrations
are small compared to the direct electrostatic coupling as well as
relevant electrical characteristics of conventional quantum dot
devices including the charging energy and energy-level spacings. In
particular, the pulse-induced vibrations would be screened in the
conductive regions. However, in the unscreened or depleted regions,
the pulse induced vibrations will be visible. Of particular
relevance, is in applications where electrical pulses are
implemented to align electronic energy levels between a lead and a
single quantum dot or between adjacent quantum dots. In such
systems, the pulse-induced vibrations could momentarily perturb the
alignment and lead to uncertainty in the evolution in the electronic
system.

Moreover, surface acoustic waves with amplitudes less than
$0.01$~mV have proved to provide an efficient means of
implementing quantized charge pumping \cite{Ebbecke:03}.

The acoustic vibrations could be described as phonons which have
been shown in experiment \cite{Fujisawa:00} and theory
\cite{Brandes:99} to couple strongly to electrons. When the energy
level separations in a quantum dot are equal to the energies of the
excited phonons, the phonons could be absorbed by electrons in the
dot. In a double quantum dot system, this could lead to inelastic
tunnelling between the adjacent dots \cite{Naber:06}.

Figure~\ref{fig:pulseFirstSim}(a) also shows the ability of the
surface acoustic wave modes to propagate for at least four microns
along the surface with no decay in amplitude. The mentioned four
microns is only a limit of the computational domain and we would
expect the induced surface acoustic wave to propagate over much
longer distances. Figure~\ref{fig:pulseFirstSim}(b) shows grayscale
plots of the induced electrical potential at time $t=2.0$~ns. The
stronger potential fields at ($x \sim 3~\mu$m) and ($x \sim
13~\mu$m) are clearly localized at the surface (i.e. $z=4~\mu$m)
with the amplitude at two wavelengths into the depth ($< 10~\mu$V)
substantially reduced compared to that at the surface ($\sim
1.0$~mV), as is characteristic of surface acoustic waves in GaAs.

\begin{figure}[!htb]
\begin{center}
\rotatebox{0}{\scalebox{0.225}[0.175]{\includegraphics*[0.1in,0.1in][12.0in,7.0in]{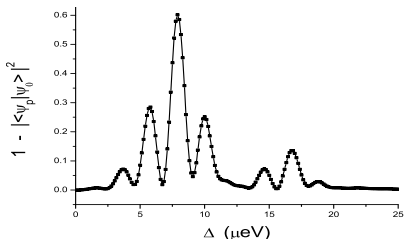}}}
\end{center}
\caption{The effect of a pulse-induced surface acoustic wave on a
double quantum dot device.} \label{fig:qubitState}
\end{figure}

In the experimental realization of quantum information processing
devices \cite{Hayashi:04,Petta:05} the applied voltage pulses are
of the order of $5$~mV. Our simulations predict the excitation of
surface acoustic waves with amplitudes of $\sim$ $10~\mu$V, which
is comparable to the inelastic tunnelling energies and the energy
level bias applied to a double dot system \cite{Hayashi:04,ft1}.
To demonstrate an effect of this pulsed-induced surface acoustic
wave on a adjacent double quantum dot system, we solve numerically
the relevant time-dependent Schr\"{o}dinger equation;

\beq  \frac{\partial }{\partial t}|\psi \rangle = -\frac{i}{2}
(\epsilon(t) \sigma_z  + \Delta(t) \sigma_x ) | {\psi} \rangle \eeq

\noindent where $\epsilon(t)$ is the energy difference between the
uncoupled localized states, $\Delta(t)$ is the energy difference
between the stationary states when the local charge states are
degenerate, and $\sigma_i$ are the Pauli matrices. In our
calculation, $\epsilon(t)$ is computed by assuming that the
quantum dot centers of the adjacent system are separated by a
distance of $200$~nm \cite{Hayashi:04}, and have a zero
energy-level bias in the absence of acoustoelectric vibrations.
This double-dot is located $5~\mu$m laterally away from the source
of surface acoustic waves. We compare the state of the neighboring
double-dot system with $|\Psi_{p}\rangle$ and without
$|\Psi_{0}\rangle$ the electric potential wave, to determine the
influence of the acoustic vibrations. Figure~\ref{fig:qubitState}
shows the infidelity between the states, as a function of $\Delta$
chosen to correspond to that in experiments (and is assumed
constant for simplicity) \cite{Hayashi:04}.

The oscillations of Fig.~\ref{fig:qubitState} approximately
commensurate with the frequency of the acoustic vibrations ($\sim
3$~GHz). The infidelity $1-|\langle \Psi_{p}|\Psi_{0}\rangle|^2$
oscillates between $0$ and $0.6$ which corresponds to a
substantial perturbation to the evolution of the quantum state.
This result is simply representative of what may happen, showing
that significant errors can occur for typical parameters and that
a judicious choice of parameters can mitigate the error.

In a system consisting of an array of quantum dots on the surface,
each of which implements electric-field pulses to control an
electronic degree of freedom, we would expect the excitation of
surface acoustic waves, which would be capable of traversing through
multiple quantum dots. The necessity to incorporate the information
regarding the electric-field pulses applied to every quantum dot,
into the description of the dynamics of each quantum dot,
potentially poses a significant challenge for the scalability of
these systems for quantum information processing applications.

In addition, the readout of the final state of a quantum system
requires the preparation, manipulation and measurement of an
ensemble of identically prepared systems and hence, the repeated
application of an electrical pulse. As a result, we could expect a
substantial accumulation of acoustoelectric vibrations within the
underlying lattice.

\begin{figure}[!htb]
\begin{center}
\epsfxsize=8.5cm \centerline{\epsffile{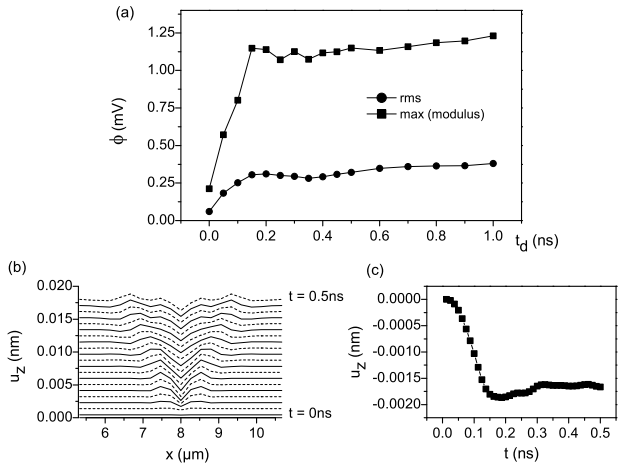}}
\end{center}
\caption{(a) The dependence of the amplitude of the pulse-induced
acoustoelectric vibrations $\phi$, on the pulse duration $t_d$.
(b) Time series plot of the $z$ component $u_z$ of displacement
across the device for a sequence of times $t=0$ to $t=0.5$~ns with
$\Delta t = 0.025$~ns. A small offset is applied between curves.
(c) For clarity, the value of $u_z$ at $x=8~\mu$m for the sequence
of times $t=0$ to $t=0.5$~ns.} \label{fig:pulseTD}
\end{figure}

The dependence of the pulse-induced acoustoelectric vibrations, on
the duration of the pulse $t_d$ is investigated by varying $t_d$
from $0$~ns to $1.0$~ns, with $t_r = 0.025$~ns and $A = 1.0$~mV.
The measurements correspond to $0.5$~ns after the end of the pulse
train. Figure~\ref{fig:pulseTD}(a) shows the computed root mean
square ($rms$) and maximum of the modulus ($max (modulus)$) of the
induced oscillations computed at $100$~nm below the surface for
data between $x=0$~nm and $x=16~\mu$m, for each value of $t_d$.
Both measures are included as the $rms$ takes into account all
wave modes (surface and bulk) whereas from the $max (modulus)$ we
see the maximum amplitude i.e. only the surface mode. One can see
a rapid increase of these measures up to $t_d \sim 0.18$~ns, after
which the $rms$ becomes flat at $\sim 0.3$~mV, while the $max
(modulus)$ measure stabilizes at $\sim 1.2$~mV. The dramatic
change in behavior at $t_d \sim 0.18$~ns, can be understood by
examining the deformation of the underlying lattice.
Figure~\ref{fig:pulseTD}(b) shows the $z$ component $u_z$ of the
lattice displacement as the voltage pulse is applied. The $x$
component exhibits similar behavior is not therefore shown. As an
external electric-field pulse is applied, the lattice slowly
deforms. After a certain amount of time $t_c$ (with the voltage
bias on), the lattice deformation reaches its equilibrium value
(for the given electric potential), and then oscillates about this
position before stabilizing. The time $t_c$ is $\sim$ $0.18$~ns
for the device modelled in this work. We would expect the time
$t_c$ to depend on the material, the device geometry, crystal
orientation with respect to the coordinate axes and the model for
the applied electric field (which in our case avoids a full
self-consistent calculation).

In our simulations, the response time $t_c$ of the lattice to the
applied electric fields is much greater than the frequently
applied rise times $t_r$ of the applied pulse which in the ideal
case is zero. Hence, for the regime of interest in semiconductor
quantum dot applications, variation of the rise time of the pulse
would have little effect on the pulsed-induced vibrations.

We may also conclude from our investigations of varying the rise
time and duration of the pulse, that the pulses implemented in
quantum information processing applications excite acoustoelectric
vibrations of the maximum amplitude (for a given amplitude for the
applied pulse).

Future work may involve the investigation of novel schemes which may
reduce the effect of the acoustoelectric vibrations. For the present
time, we point out that surface acoustic waves may be utilized for
quantum information processing applications \cite{Barnes:00}.
Moreover, materials which have a spatial inversion center such as Si
are immune to the excitation of acoustoelectric vibrations through
electric field pulses and thus have a significant advantage for
quantum control and quantum computation. Devices based on such
materials are currently being developed for quantum computation
\cite{Gorman:05,Rahman:05:3,Kane:98}.

Many spin-readout proposals rely on mapping a spin state to a charge
state, since the detectors that are likely to be used are Quantum
Point Contacts (QPC's). Therefore, the results are also relevant to
spin qubit systems \cite{Barrett:06,Stace:04}.

\section*{CONCLUSION}

In conclusion, we have demonstrated the excitation of
acoustoelectric modes of vibration in a GaAs lattice as a result of
sharp nano-second electric-field pulses applied to surface gates.
The excited wave modes include surface acoustic modes which have
been shown to travel at least four microns along the free surface of
a material. We have demonstrated the substantial perturbation these
waves may have on the evolution of quantum systems. Whilst the
effect we have considered does not decohere the system in the normal
sense of becoming correlated with the environment, the fact the
acoustoelectric waves induced by gates at one place can
significantly affect the state of a remote double-dot system, means
that the results are very significant for quantum information
processing applications.

\section*{ACKNOWLEDGEMENTS}

We thank T. Fujisawa, T. Hayashi, C. M. Marcus, C. J. B. Ford, J.
Jefferson, R. Young and S. Pfaendler for comments, useful
discussions and assistance.  SR acknowledges the Cambridge-MIT
Institute for financial support.


\begin{references}
\bibitem{ft0} When air is considered above the substrate, the
boundary condition consists of the continuity of $D_z$. The
present boundary condition is useful when air is neglected for
computational speed.
\bibitem{ft1} We have performed simulations which suggest that variation of the gate
width within the quantum dot regime, does not have a significant
effect on the amplitude of the induced vibrations.
\end{references}
\end{document}